\documentclass[3p,times]{elsarticle}

\usepackage{ecrc}
\usepackage{amssymb,amsfonts,amsmath}

\volume{00}
\def\lsim{\mathrel{\rlap{\lower3pt\hbox{\hskip1pt$\sim$}}
     \raise1pt\hbox{$<$}}} 
\def\gsim{\mathrel{\rlap{\lower3pt\hbox{\hskip1pt$\sim$}}
     \raise1pt\hbox{$>$}}} 
\newcommand{\beq}{\begin{equation}}
\newcommand{\eeq}{\end{equation}}
\newcommand{\bea}{\begin{eqnarray}}
\newcommand{\eea}{\end{eqnarray}}
\firstpage{1}

\journalname{Nuclear Physics A}

\runauth{}

\jid{nupha}

\usepackage{amssymb}

\usepackage[figuresright]{rotating}

\begin{document}

\begin{frontmatter}



\dochead{}

\title{Recent results on QCD thermodynamics: lattice QCD versus Hadron Resonance Gas model}


\author[1]{Szabolcs~Bors\'{a}nyi}
\author[1,2,3]{Zolt\'{a}n Fodor}
\author[1]{Christian Hoelbling}
\author[3]{S\'{a}ndor~D.~Katz}
\author[1,4]{Stefan~Krieg}
\author[1,5]{Claudia Ratti}
\author[1]{K\'alm\'an~K.~Szab\'o}
\address[1]{Department of Physics, University of Wuppertal, Gau\ss str. 20, D-42119, 
Germany}
\address[2]{Forschungszentrum J\"ulich, J\"ulich, D-52425, Germany}
\address[3]{Institute for Theoretical Physics, E\"otv\"os University, P\'azm\'any
1, H-1117 Budapest, Hungary}
\address[4]{Center for Theoretical Physics, MIT, Cambridge, MA 02139-4307, USA }
\address[5]{Dipartimento di Fisica Teorica, Universit\'a degli Studi di Torino, via Giuria 1, 10125 Torino, Italy}
\begin{abstract}
We present our most recent investigations on the QCD cross-over transition
temperatures with 2+1 staggered flavours and one-link stout improvement [JHEP 1009:073, 2010].  We
extend our previous two studies [Phys. Lett. B643 (2006) 46, JHEP 0906:088
(2009)] by choosing  even finer lattices ($N_t$=16) and we work again with
physical quark masses. All these results are confronted with the
predictions of the Hadron Resonance Gas model and Chiral Perturbation Theory
for temperatures below the transition region. Our results can be reproduced by
using the physical spectrum in these analytic calculations. A comparison with the results of the
hotQCD collaboration is also discussed.
\end{abstract}




\end{frontmatter}


\section{ Introduction}
\label{intro}
One of the most interesting quantities that can be extracted from lattice
simulations is the transition temperature $T_c$ at which hadronic matter is
supposed to undergo a transition to a deconfined, quark-gluon phase. This
quantity has been vastly debated over the last few years, due to the
disagreement on its numerical value observed by different lattice
collaborations, which in some cases is as high as 20\% of the absolute value.

Indeed, the analysis of the hotQCD collaboration (performed with two different
improved staggered fermion actions, asqtad and p4, and with physical strange
quark mass and somewhat larger than physical $u$ and $d$ quark masses, $m_s
/m_{u,d} = 10$), indicates that the transition region lies in the range $T =
(185-195)$ MeV
(for the latest published result and for references see \cite{Bazavov:2009zn}).
The results obtained
by our collaboration using the staggered stout action (with physical light and
strange quark masses, thus $m_s/m_{u,d}\simeq28$) show that the
value of the transition temperature lies in the range 150-170 MeV, and it
changes with the observable used to define it \cite{6, 7}. This reflects the cross-over 
nature of the transition \cite{8}.
We present our most recent results for several physical quantities (for all details see \cite{Borsanyi:2010bp}): our previous
works \cite{6, 7} have been extended to an even smaller lattice spacing
(down to $a \lsim 0.075$ fm at $T_c$), corresponding to $N_t$=16. 
We use physical light and strange quark masses: we fix them by reproducing
$f_K/m_\pi$ and $f_K/m_K$ and by this procedure \cite{7} we get $m_s /m_{u,d} = 28.15$.
We also present some aspects of the Hadron Resonance Gas 
model and the comparison 
between HRG model results and the lattice data from ours and the hotQCD collaborations\footnote{Note, that
recently preliminary results were presented \cite{Bazavov:2010sb} and
the results of the hotQCD collaboration moved closer to our results.}. 
As we will see, our analysis provides a straightforward explanation for the observed discrepancy
in the results of the two collaborations.


\section{Details of the lattice simulations}
We use \cite{6,7} a tree-level Symanzik 
improved gauge, and a stout-improved staggered fermionic action (see Ref. \cite{Aoki:2005vt} for details).  
The stout-smearing is an important part of the framework, which  
reduces the taste violation. Indeed, in \cite{Fodor:2007sy} we pointed out that the continuum limit can be approached only if one reduces the unphysical pion splitting (the main
motivation of our choice of action).

\begin{figure}
\hspace{.8cm}
\begin{minipage}{.48\textwidth}
\parbox{6cm}{
\scalebox{.55}{
\includegraphics{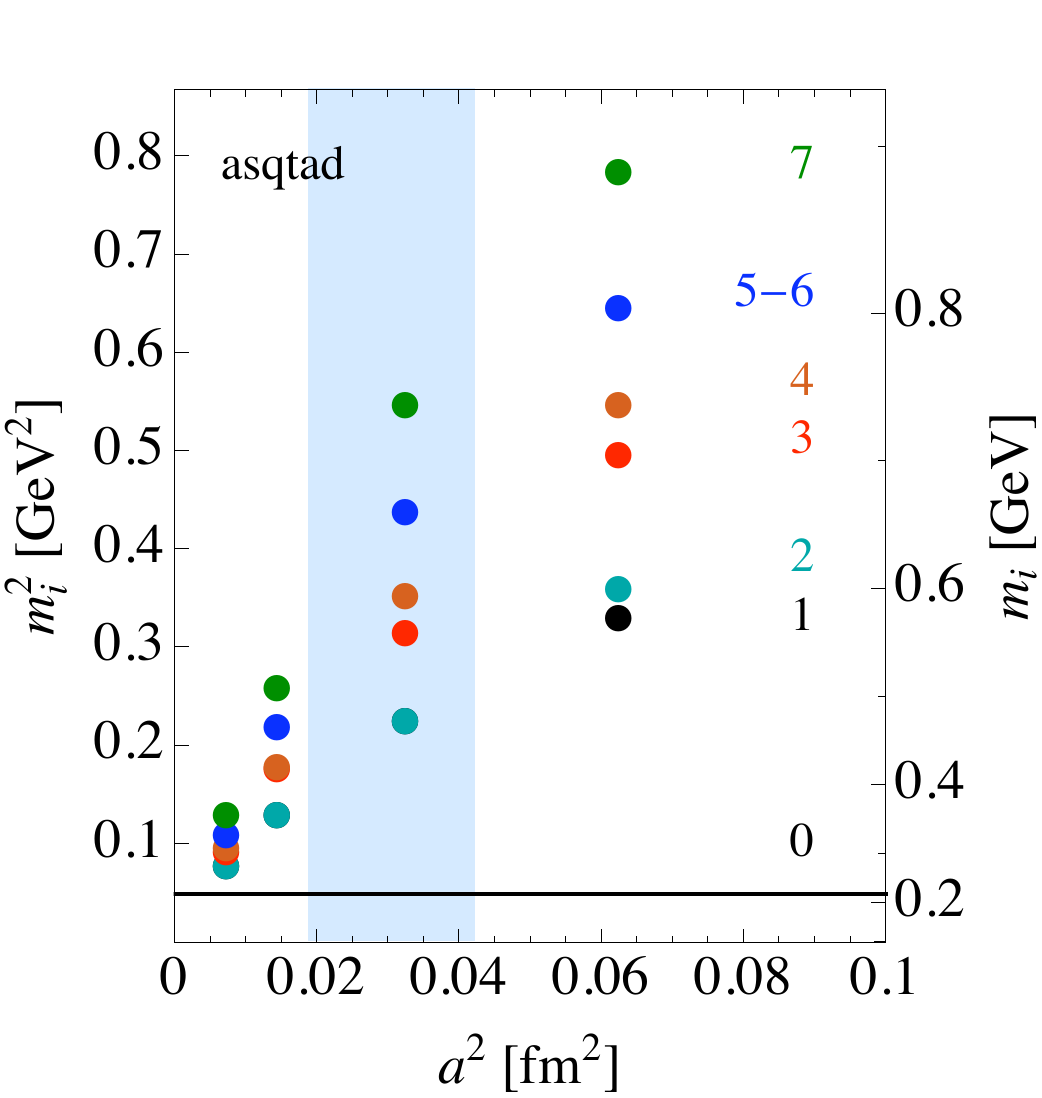}\\}}
\end{minipage}
\hspace{.4cm}
\begin{minipage}{.48\textwidth}
\hspace{.4cm}
\parbox{6cm}{
\scalebox{.55}{
\includegraphics{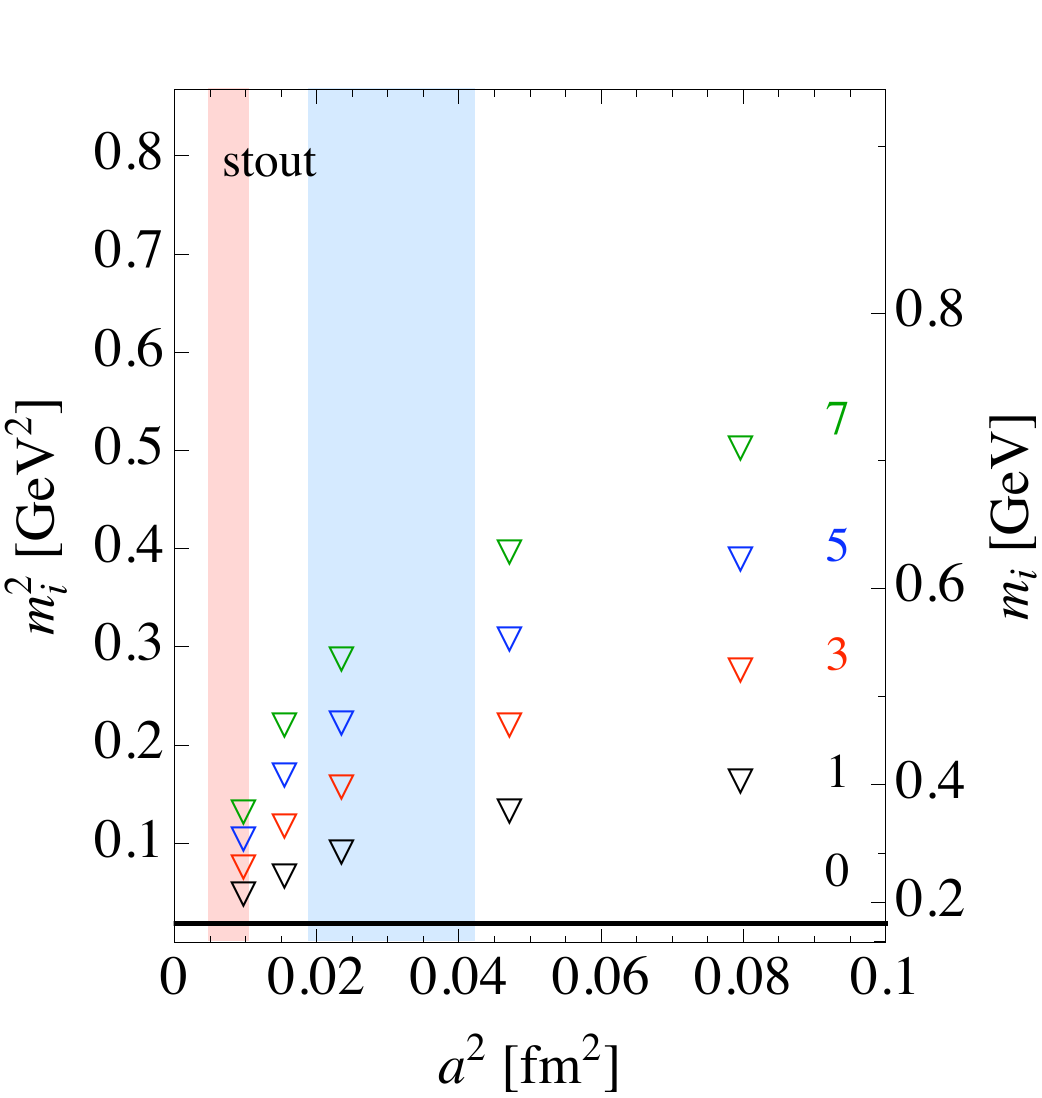}\\}}
\end{minipage}
\caption{Masses of the pion multiplet squared, as functions of the lattice spacing squared. 
Left panel: asqtad action \cite{Bazavov:2009bb}. Right panel: stout action. The numbers
next to the data correspond to the taste matrices (see Ref. \cite{Borsanyi:2010bp} for details). In both panels, the blue band indicates the relevant range of
lattice spacings for a thermodynamics study at $N_t=8$ between $T=$ 120 and 180
MeV. The red band in the right panel corresponds to the same temperature range
and $N_t=16$. In both figures, the horizontal line labelled as ``0" is the
pseudo-Goldstone boson, which has a mass of 220 MeV for the asqtad results, and
135 MeV for the stout ones 
}
\label{fig1}
\end{figure}

The pion splitting of a staggered framework 
has to vanish in the continuum limit. Once it 
shows an $\alpha_s a^2$ dependence (in practice $a^2$ dependence
with a subdominant logarithmic correction) we are in the scaling region.  
This is an important 
check for the validity of the staggered framework at a given
lattice spacing. 
In Fig. \ref{fig1} we show the leading order $a^2$-behavior of the 
masses of the pion multiplets calculated with the asqtad (left) and stout (righ) actions. It is evident that the 
continuum expectation is reached faster in the stout action than in the asqtad one. In addition, 
in the present 
paper we push our results to $N_t = 16$, which corresponds to even smaller lattice 
spacings and mass splittings than those used in \cite{7}. 

In analogy with what we did in \cite{6,7}, we set the scale at the physical 
point by simulating at $T=0$ with physical quark masses \cite{7} and reproducing the kaon and pion masses and the kaon decay constant. This gives 
an  uncertainty of about 2\% in the scale setting, which propagates in the uncertainty in the 
determination of the temperature values.

\section{Lattice vs Hadron Resonance Gas model results}
We present our lattice results for the strange quark number
susceptibility, chiral
condensate and interaction measure. We extract the values
of the transition temperature associated to these observables. 
We also perform a HRG analysis and compare the HRG predictions to our results and to those of the 
hotQCD Collaboration. 

The Hadron Resonance Gas model has been widely used to study the
low temperature phase of QCD in comparison with lattice data.
In Ref.
\cite{Huovinen:2009yb} an important ingredient was included in this model,
namely the pion mass- and lattice spacing-dependence of the hadron masses. This gives
rise to a resonance spectrum which is distorted by lattice artifacts, and which
needs to be taken into account in the comparison with the results of the hotQCD collaboration. Here
we combine these ingredients with Chiral Perturbation Theory
($\chi$PT)  \cite{14}. This opens the possibility to study chiral quantities,
too.
The HRG model has its roots in the theorem by Dashen, Ma and Bernstein \cite{15}, 
which allows to calculate the microcanonical partition function of an interacting
system, in the thermodynamic limit $V\rightarrow\infty$, 
to a good approximation, assuming that it is a gas of non-interacting free hadrons and resonances \cite{16}. 
The pressure of the Hadron Resonance Gas model can be written as 
the sum of independent contributions coming from non-interacting resonances.
We include all known baryons and mesons up to 2.5 GeV, as listed in the latest edition of the Particle 
Data Book (for an improvement of the model by including an exponential mass spectrum see
\cite{NoronhaHostler:2007jf}). 
We will compare the results obtained with the physical hadron masses to those 
obtained with the distorted hadron spectrum which takes into account lattice discretization effects. 
Each pseudoscalar meson in the staggered formulation is split into 16 mesons with different masses, which
are all included.
Similarly to Ref. \cite{Huovinen:2009yb}, we will also take into account the pion mass- and lattice spacing- 
dependence of all other hadrons and resonances. 

\begin{figure}
\hspace{.8cm}
\begin{minipage}{.48\textwidth}
\parbox{6cm}{
\scalebox{.6}{
\includegraphics{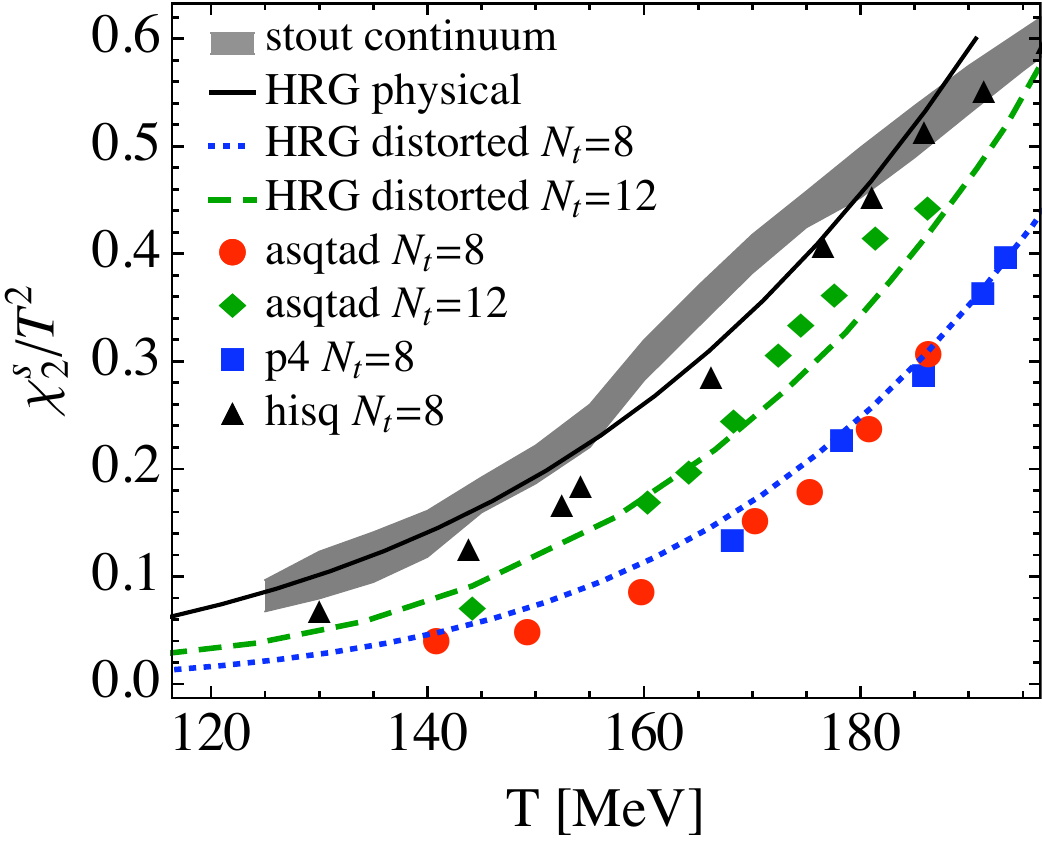}\\}}
\end{minipage}
\hspace{.4cm}
\begin{minipage}{.48\textwidth}
\hspace{.4cm}
\parbox{6cm}{
\scalebox{.56}{
\includegraphics{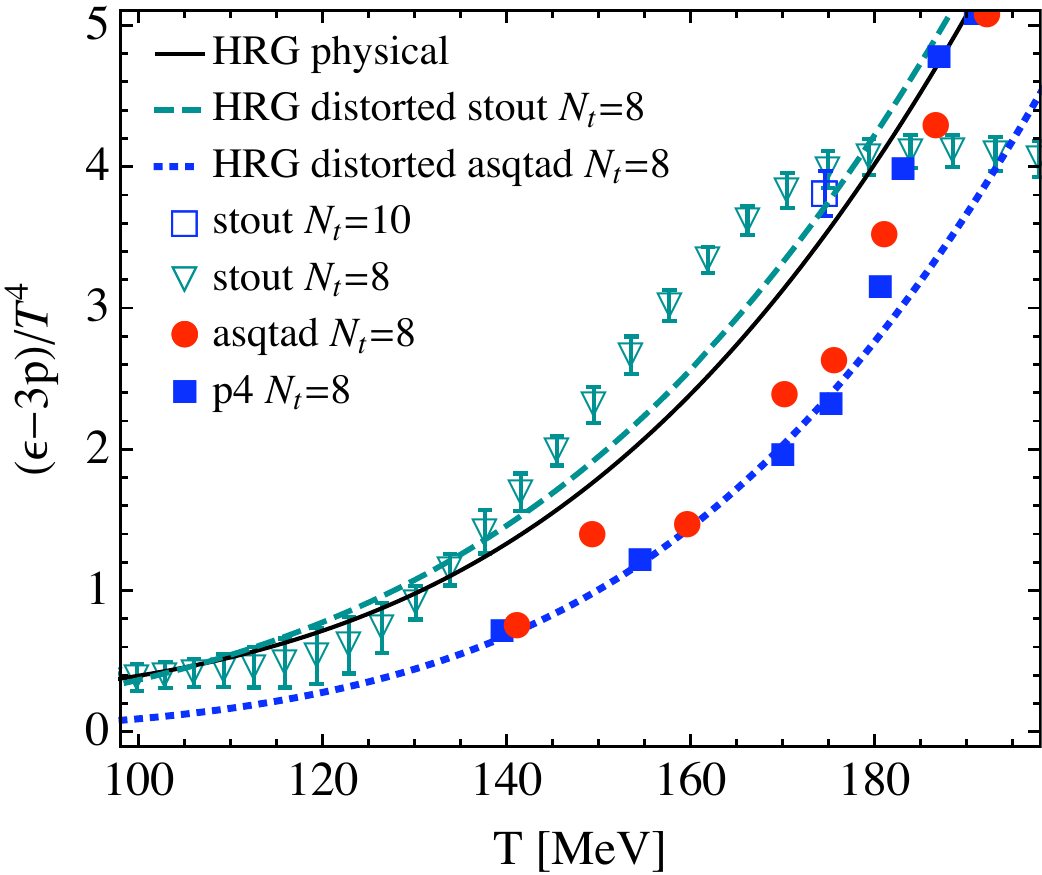}\\}}
\end{minipage}
\caption{
Left panel: strange quark susceptibility as a function of the temperature. 
full symbols correspond to results obtained with the asqtad, p4 and hisq
actions \cite{Bazavov:2009zn,Bazavov:2010sb}. Our continuum result is indicated by the gray band. The solid line is the HRG model result with physical
masses. The dashed and dotted lines are the
HRG model results with distorted masses corresponding to $N_t=12$ and $N_t=8$, which take into account the discretization
effects and heavier quark masses, which characterize the results of the hotQCD Collaboration.
Right panel: $(\epsilon-3p)/T^4$ as a function of the temperature.
Open symbols are our results. Full symbols are the results for the asqtad and p4 actions at $N_t=8$ \cite{Bazavov:2009zn}. Solid line: HRG model with physical masses. Dashed lines: HRG model with distorted spectrums. As it can be seen, the prediction of the HRG model with a spectrum distortion corresponding to the stout action
at $N_t=8$ is already quite close to the physical one. The error on the recent preliminary HISQ result \cite{Bazavov:2010sb} is larger than the difference between the stout and asqtad data, that is why we do not show them here.
}
\label{fig2}
\end{figure}

Quark number susceptibilities 
increase during the transition, therefore
they can be used to identify this region. They are defined as
$\chi_{2}^{q}=\left.\frac{T}{V}\frac{\partial^2\ln Z}{\partial(\mu_q)^2}\right|_{\mu_i=0},$
 (with $q=u,d,s$).
In the left panel of Fig. \ref{fig2} we show our continuum-extrapolated results for the strange quark
number susceptibility, in comparison with the HRG results with physical spectrum. Also shown are the hotQCD collaboration data, in comparison with the HRG model results with distorted spectrum.
In the right panel of Fig. \ref{fig2} we show the trace anomaly ($\epsilon-3p$) divided by $T^4$ as a function of
the temperature. Our $N_t=8$ results are taken from Ref. \cite{EOS}. Notice
that, for this observable, we have a check-point at $N_t=10$: the results
are on top of each other. Also shown are the results
of the hotQCD collaboration at $N_t=8$ \cite{Bazavov:2009zn} and the HRG model predictions for physical and distorted resonance spectrums. On the one hand, our results are in good agreement with the ``physical'' HRG model ones. It is important to note, that using our mass splittings and inserting this  distorted spectrum into the HRG model gives a temperature dependence which lies essentially on the physical HRG curve (at least within our accuracy). On the other hand, a distorted spectrum based on the asqtad and p4 frameworks results in a shift of about 20 MeV to the right. 

\begin{figure}
\hspace{.8cm}
\begin{minipage}{.48\textwidth}
\parbox{6cm}{
\scalebox{.56}{
\includegraphics{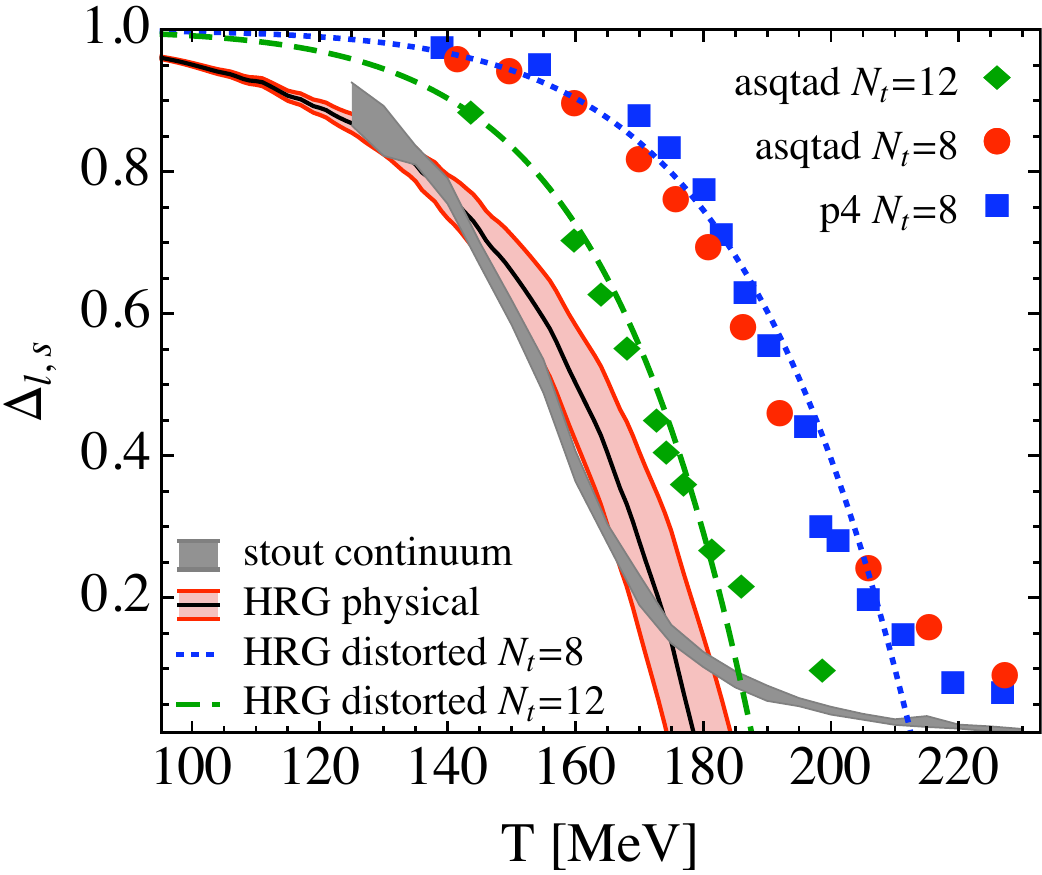}\\}}
\end{minipage}
\hspace{.4cm}
\begin{minipage}{.48\textwidth}
\hspace{.4cm}
\parbox{6cm}{
\scalebox{.56}{
\includegraphics{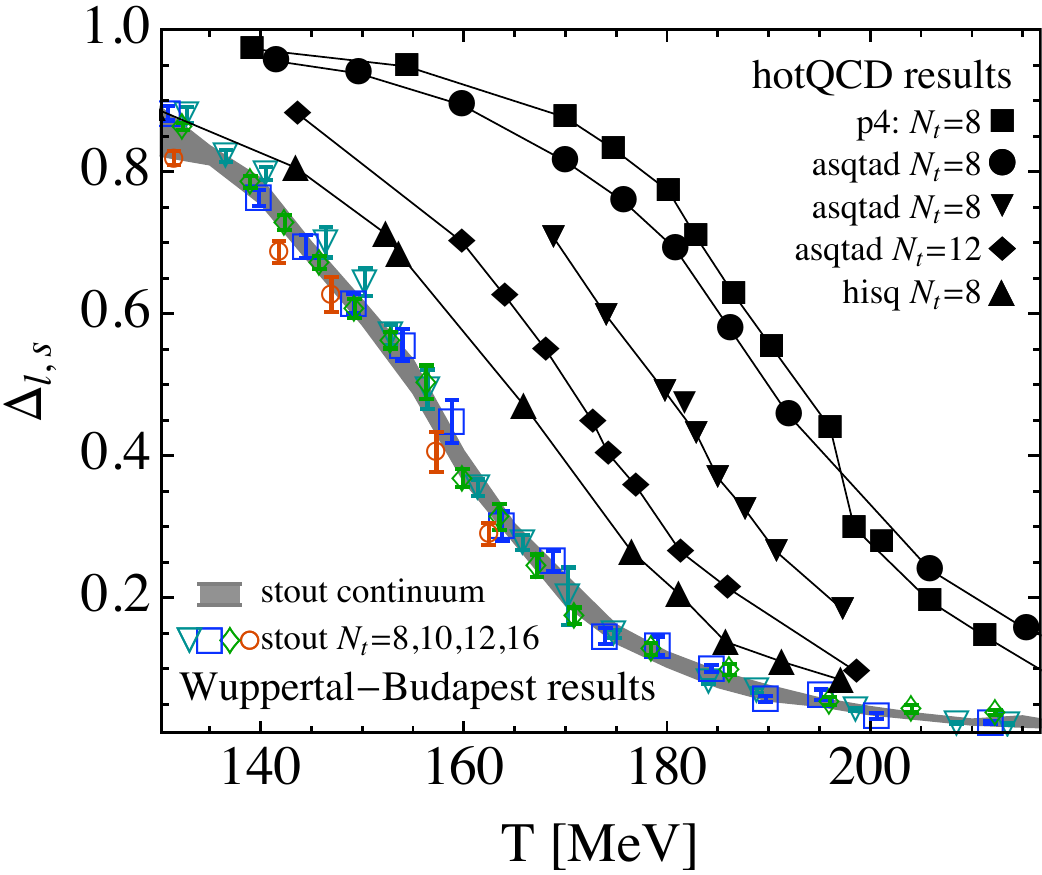}\\}}
\end{minipage}
\caption{
Subtracted chiral condensate $\Delta_{l,s}$ as a function of the 
temperature. Gray bands are the continuum results of our collaboration, obtained with the 
stout action. Full symbols are obtained with the asqtad 
and p4 actions \cite{Bazavov:2009zn,Bazavov:2010sb}. In the left panel, the solid line is the HRG model+$\chi$PT result with physical masses. 
The error band corresponds to the uncertainty in the quark mass-dependence of 
hadron masses. The dashed lines are the HRG+$\chi$PT model results with distorted masses, 
which take into account the discretization effects and heavier quark masses used in 
\cite{Bazavov:2009zn,Bazavov:2010sb} for $N_t=8$ and $N_t=12$.
In the right panel we show a comparison between stout, asqtad, p4 and
HISQ results. Our results are shown by
colored open symbols, whereas the hotQCD results are shown by full black
symbols. The gray band is our continuum result, the thin lines for the hotQCD
data are intended to lead the eye. Our stout results were all obtained by the
physical pion mass of 135 MeV. The full dots and squares correspond to
$m_\pi=220$ MeV, the full triangles and diamonds correspond to $m_\pi=160$ MeV
of the hotQCD collaboration. 
}
\label{fig3}
\end{figure}
In order to compare our results to those of the hotQCD collaboration, we also calculate the quantity
$\Delta_{l,s}=(\langle\bar{\psi}\psi\rangle_{l,T}-\frac{m_l}{m_s}\langle\bar{\psi}\psi\rangle_{s,T})/
\langle\bar{\psi}\psi\rangle_{l,0}-\frac{m_l}{m_s}\langle\bar{\psi}\psi\rangle_{s,0})$ (with $l=u,d$).
We compare our results to the predictions of the HRG model and $\chi$PT
\cite{18}. To this purpose, we need to know the quark mass dependence of the masses of all
resonances included in the partition function. We assume that all resonances behave as their fundamental states as functions of the quark mass, and take this information from Ref. \cite{MartinCamalich:2010fp}. They agree with the results obtained by our collaboration in \cite{19}.

\section{Conclusions}
We have presented our latest results for the QCD transition temperature. The 
quantities that we have studied are the strange quark number susceptibility,
the chiral condensate and the trace anomaly. We have given the complete
temperature dependence of these quantities, which provide more
information than the characteristic temperature values alone.
Our previous results have been pushed to an even finer lattice ($N_t = 16$).
The new data corresponding 
to $N_t = 16$ confirm our previous results. The trace anomaly \cite{EOS} was obtained for $N_t=8$ and a check-point at $N_t=10$. The transition temperature that we obtain from this last quantity is very close to the one obtained from the chiral condensate. 
In order to find the origin of the discrepancy between the results of our collaboration and the hotQCD ones, we calculated these observables in the Hadron 
Resonance Gas model. Besides using the physical hadron masses, we also
performed the calculation with modified masses which take into account the
heavier pions and larger lattice spacings used in \cite{Bazavov:2009zn}. We
find an agreement between our data and the HRG ones with ``physical'' masses,
while the hotQCD collaboration results are in agreement with the HRG model once
the spectrum is ``distorted'' as it was directly measured on the lattice
\cite{Bazavov:2009bb}.
All the details can be found in Ref. \cite{Borsanyi:2010bp}.
\bibliographystyle{elsarticle-num}



\end{document}